\documentclass[article]{revtex4}
\usepackage{graphicx}
\usepackage{amssymb}
\usepackage{bm}
\usepackage{hyperref}
\usepackage{epstopdf}
\usepackage{color}
\usepackage{mathrsfs}
\begin{document}
\title{Diagnostics and future evolution analysis of the two parametric models}

\author{Guang Yang$^{1}$}
\email{yang-guang@mail.nankai.edu.cn}
\author{Deng Wang$^{2}$}
\email{Cstar@mail.nankai.edu.cn}
\author{Xinhe Meng$^{1,3}$}
\email{xhm@nankai.edu.cn}

\affiliation{
$^1${Department of Physics, Nankai University, Tianjin 300071, China}\\
$^2$Theoretical Physics Division, Chern Institute of Mathematics, Nankai University, Tianjin 300071, China\\
$^3$State Key Laboratory of Theoretical Physics China,CAS, Beijing 100190, China\\}

\begin{abstract}
In this paper, we apply three diagnostics including $Om$, Statefinder hierarchy and the growth rate of perturbations into discriminating the two parametric models for the effective pressure with the $\Lambda$CDM model. By using the $Om$ diagnostic, we find that both the model 1 and the model 2 can be hardly distinguished from each other as well as the $\Lambda$CDM model in terms of 68\% confidence level. As a supplement, by using the Statefinder  hierarchy diagnostics and the growth rate of perturbations, we discover that not only can our two parametric models be well distinguished from $\Lambda$CDM model, but also, by comparing with $Om$ diagnostic, the model 1 and the model 2 can be distinguished better from each other. In addition, we also explore the fate of universe evolution of our two models by means of the rip analysis.
\end{abstract}

\maketitle
\section{Introduction}
Since the Type Ia supernovae observation in 1998 indicates that the current universe is in the phase of accelerated expansion, increasing amounts of observations verify this surprisingly exotic phenomenon \cite{a1,a2}. In order to accommodate this phenomenon, various models as dark energy are proposed. The basic idea of dark energy comes up in the context of supposing the general theory of relativity works precisely well at the cosmological scales, a perfect fluid with effectively large enough negative pressure is required to accelerate the universe expansion. According to the Planck's result \cite{0}, vacuum occupies about $70\%$ of the total energy density of our universe. Up to now, many attempts have been made to understand the dynamics of dark energy, among which a spatially flat $\Lambda$CDM cosmology model specified by six parameters (we will refer to as the basic model) gets the best accordance with the astronomical observations though suffering from some serious fundamental physics problems. It is still worth mentioning that recent Planck results is statistically consistent with the basic assumption of the $\Lambda$CDM model. But theoretical difficulties in the $\Lambda$CDM model is not alleviated by the newest observations. Based on this concern, we proposed two parametric models for the total pressure to explore the universe at the late-time evolution stage in our previous work~\cite{manuscript}. In order to understand the evolutionary history of the two models more clearly, in this paper we are aimed at discriminating these two models and $\Lambda$CDM model by using the Om diagnostic~\cite{state19}, the Statefinder hierarchy~\cite{state1,state1',state17} and the growth rate of perturbations \cite{1,2}. Moreover, we also investigate the fates of universe of these two models via the rip analysis and find the model 2 have the possibility that the universe will terminate after approximate $10^{11}$ years.

This paper is organized as follows: in section II, we briefly review our two parametric models---model 1 and model 2. In Section III, we review the $Om$ diagnostic, Statefinder hierarchy and the growth rate of perturbations, and use these three geometrical diagnostics to distinguish the model 1 and the model 2 from the $\Lambda$CDM model. In Section IV, we discuss the fate of universe evolution of our two models by means of the rip analysis. In Section V, We end the paper with discussions and conclusions.

\section{A brief review of two parametric models for the total pressure}

{\bf Model 1}: model 1 is written as
\begin{equation}
  P(z)=P_a+P_b z \label{eq2-1},
\end{equation}
\noindent
where $P_a$, $P_b$ are model parameters.

Consider the relation between the scale factor $a$ and the redshift $z$:
\begin{equation}
  a=\frac{a_0}{1+z}=\frac{1}{1+z}\label{eq2-2},
\end{equation}

\noindent
where $z$ is the redshift. Substitute Eqs.~(\ref{eq2-1})and~(\ref{eq2-2}) to energy conservation equation $\dot{\rm \rho}+3 H(P+\rho)=0$, we can obtain the expression of the total energy density
\begin{equation}
\rho(a)=-(P_a-P_b)-\frac{3}{2}P_b a^{-1}+C_1 a^{-3} \label{eq2-3},
\end{equation}

\noindent
where $C_1$ is an integration constant and $C_1 a^{-3}$ corresponds to the dust matter, set the total energy density $\rho_0$ today, then we have $C_1=\rho_0+P_a+\frac{1}{2}P_b$. Following ~\cite{manuscript}, rewrite the expressions of total density Eq.~(\ref{eq2-3}) and total pressure Eq.~(\ref{eq2-1})
\begin{eqnarray}
  \rho(a)&=&\rho_0[\alpha+\beta a^{-1}+(1-\alpha-\beta) a^{-3}] \label{eq2-4}, \\
  P(a)&=&\rho_0(-\alpha-\frac{2}{3}\beta a^{-1}) \label{eq2-5}.
\end{eqnarray}

\noindent
Here the parameters $(P_a,P_b)$ are replaced by the new dimensionless parameters $(\alpha,\beta)$, where $\alpha\equiv -\frac{P_a-P_b}{\rho_0}$ and $\beta\equiv -\frac{3}{2}\frac{P_b}{\rho_0}$. The equation of state (EoS) of dark energy and the dimensionless Hubble parameter are given by, respectively
\begin{eqnarray}
  \omega_{de}&=&-1+\frac{\frac{1}{3}\beta(1+z)}{\alpha+\beta(1+z)} \label{eq2-6}, \\
  E(z)&=&\frac{H(z)}{H_0}=[\alpha+\beta (1+z)+(1-\alpha-\beta)(1+z)^3]^\frac{1}{2} \label{eq2-7}.
\end{eqnarray}

From Eq.~(\ref{eq2-6}), we can see in the scenario of quintessence $\beta>0$, and that in the scenario of phantom $\beta<0$. The best-fit values of parameters we take are: $\alpha=0.770931$, $\beta=-0.057783$.

{\bf Model 2}: model 2 is given by
\begin{equation}
  P(z)=P_c+\frac{P_d}{1+z} \label{eq2-8},
\end{equation}
\noindent
where $P_c$, $P_d$ are model parameters. Substitute Eqs.~(\ref{eq2-2}) and ~(\ref{eq2-8}) to energy conservation equation, one can obtain the expression for the total energy density for model 2:
\begin{equation}
\rho(a)=-P_c-\frac{3}{4}P_d a+C_2 a^{-3} \label{eq2-9},
\end{equation}

\noindent
where $C_2$ is the integration constant and $C_2 a^{-3}$ corresponds to dust matter. Set the present energy density $\rho_0$, we have $C_2=\rho_0+P_c+\frac{3}{4}P_d$. Replace the parameters $(P_c,P_d)$ with the new dimensionless parameters $(\gamma,\delta)$, then the expressions of the total density Eq.~(\ref{eq2-9}) and total pressure Eq.~(\ref{eq2-8}) can be rewritten as
\begin{eqnarray}
  \rho(a)&=&\rho_0[\gamma+\delta a+(1-\gamma-\delta) a^{-3}] \label{eq2-10}, \\
  P(a)&=&\rho_0(-\gamma-\frac{4}{3}\delta a) \label{eq2-11},
\end{eqnarray}

\noindent
where $\gamma\equiv -\frac{P_c}{\rho_0}$ and $\delta\equiv -\frac{3}{4}\frac{P_d}{\rho_0}$. The EoS of dark energy and the dimensionless Hubble parameter take the form, respectively
\begin{eqnarray}
  \omega_{de}&=&-1-\frac{\frac{1}{3}\delta (1+z)^{-1}}{\gamma+\delta (1+z)^{-1}} \label{eq2-12}, \\
  E(z)&=&[\gamma+\delta (1+z)^{-1}+(1-\gamma-\delta)(1+z)^3]^\frac{1}{2} \label{eq2-13}.
\end{eqnarray}

From Eq.~(\ref{eq2-12}), we can see in the scenario of quintessence $\delta<0$, and that in the scenario of phantom $\delta>0$. The best-fit values of parameters we take are: $\gamma=0.634812$, $\delta=0.078687$.

\section{Discriminations by the Statefinder hierarchy and $Om(z)$ diagnostic}
\subsection{$Om$ diagnostic}
The $Om$ diagnostic is an effective method to distinguish dark energy models~\cite{state19}, which has been used to distinguish $\Lambda$CDM with quintessence~\cite{state19}, phantom~\cite{state19}, PKK~\cite{state10}, holographic dark energy~\cite{state11} and SRDE~\cite{state16} in the literature. $Om$ is defined as follows
\begin{equation}
Om(x)=\frac{E^2(x)-1}{x^3-1} \label{eqs-23},
\end{equation}

\noindent
where $x=1+z$, $E(x)=\frac{H(x)}{H_0}$. Ignoring the radiation at low redshift, for $\Lambda$CDM, we have
\begin{equation}
E^2(x)=\Omega_{m0}x^3+(1-\Omega_{m0})  \label{eqs-24}.
\end{equation}

Substitute Eq.~(\ref{eqs-24}) to Eq.~(\ref{eqs-23}), we have
\begin{equation}
Om(x)\mid_{\Lambda CDM}=\Omega_{m0} \label{eqs-25}.
\end{equation}

From Eq.~(\ref{eqs-25}), we can see that $Om$ diagnostic provides a null test of $\Lambda$CDM, as for other dark energy models $Om$ is expected to give different results. Next, we will use $Om$ diagnostic to distinguish our two models from $\Lambda$CDM.

\begin{figure}[h!]
\begin{center}
\includegraphics[width=0.5\textwidth]{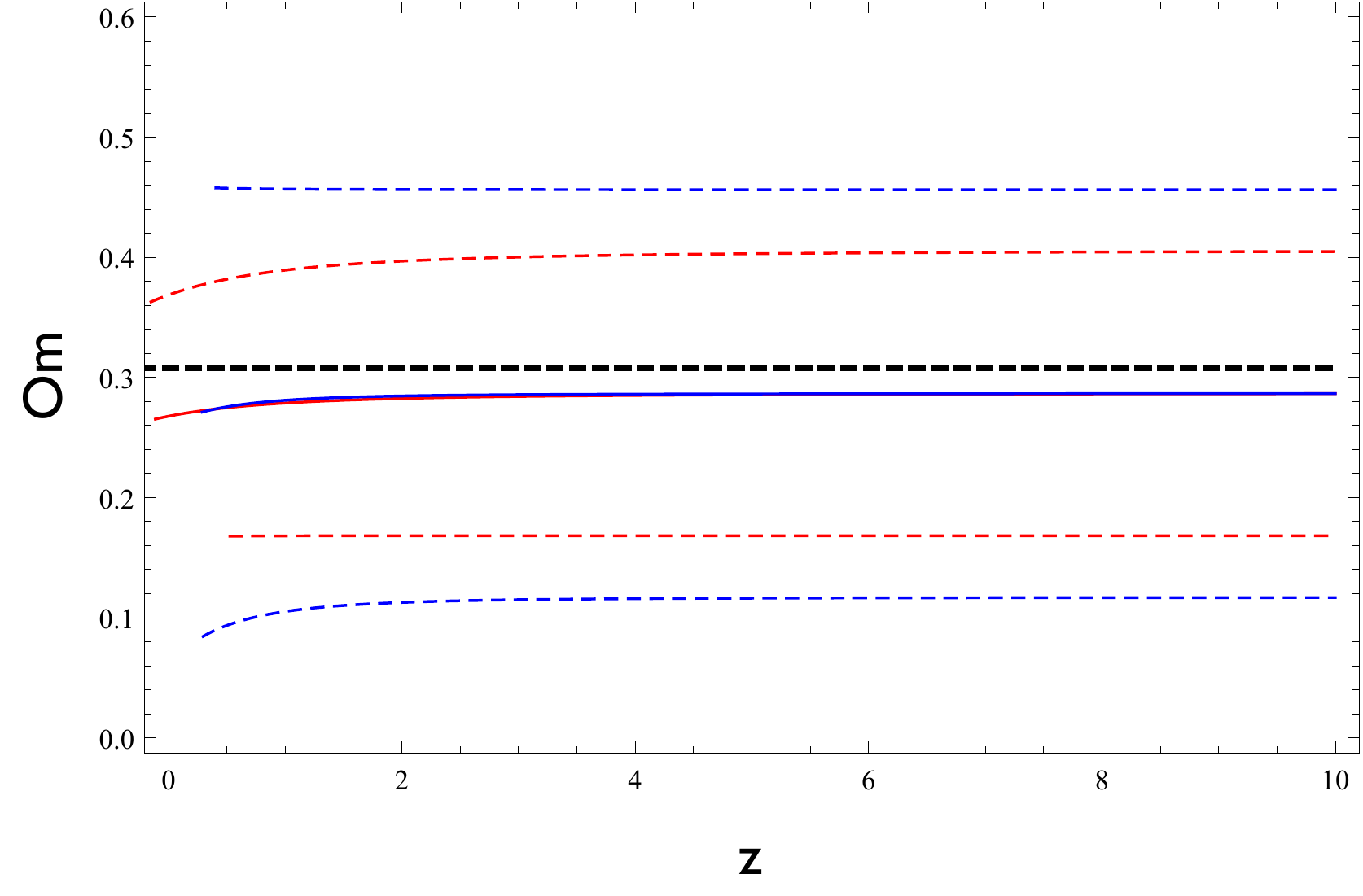}
\end{center}
\caption{The $Om$ diagnostic is shown for model 1, model 2 and $\Lambda$CDM model. The solid lines (red and blue) correspond to the best fitting values of model 1 and model 2. The red dashed lines and blue dashed lines represent the 68\% confidence level of model 1 and model 2, respectively. The horizontal black dashed line corresponds to the $\Lambda$CDM model.}
\label{fig:Om}
\end{figure}

In Fig.~(\ref{fig:Om}), we plot the evolutional trajectories of our two models and $\Lambda$CDM model. We can see that, the trajectories of the best fitting values of model 1 and model 2 can be distinguished from $\Lambda$CDM model very well at low redshift, but they can not be distinguished from $\Lambda$CDM at high redshift where the matter dominates, in addition, model 1 and model 2 can be hardly distinguished from each other in terms of 68\% confidence level. Therefore, we will adopt  Statefinder hierarchy and the growth rate of perturbations in order to discriminate the two models better.
\subsection{Statefinder Hierarchy}
With the increasing of dark energy models, how to discriminate various dark energy models becomes more and more important and meaningful. So, given this, Sahni et al.~\cite{state1} proposed a geometrical diagnostic called statefinder defined by geometric parameters $(r,s)$ to discriminate different dark energy models, such as $\Lambda$CDM, quintessence~\cite{state2,state3,state4}, GSG~\cite{state4,state5,state6}, DGP~\cite{state7,state8}, Galileon-modified gravity~\cite{state8,state9}, purely kinetic k-essence model(PKK)~\cite{state10},holographic dark energy~\cite{state11,state12}, Ricci Dark Energy model~\cite{state13}, Agegraphic Dark Energy Model~\cite{state14}, quintom dark energy model~\cite{state15} and spatial Ricci scalar dark energy(SRDE)~\cite{state16}.The geometric parameters $(r,s)$ are defined as~\cite{state1,state1'}
\begin{eqnarray}
  r&\equiv&\frac{\dddot{a}}{aH^3} \label{eqs-1} ,\\
  s&\equiv&\frac{r-1}{3(q-\frac{1}{2})} \label{eqs-2},
\end{eqnarray}

where $q\equiv-\frac{\ddot{a}}{aH^2}$ is the deceleration parameter. From Eqs.~(\ref{eqs-1}) and ~(\ref{eqs-2}), we can see parameters $(r,s)$ are just associated with the scale factor $a$ and its higher derivatives. Different dark energy models correspond to different trajectories in r-s plain, $\Lambda$CDM model corresponds to the fixed point $(r,s)=(1,0)$. Later Arabaslmani and Sahni improved a new diagnostic called ``statefinder hierarchy'' based on statefinders, all members of statefinder hierarchy can be expressed as functions of the deceleration parameter $q$ or the matter energy density parameter $\Omega_m$. Around the present time $t_0$ through Taylor expansion, the scale factor $a(t)$ can be written as
\begin{equation}
 \frac{a(t)}{a_0}=1+\sum\limits_{n=0}^\infty\frac{A_n(t_0)}{n!}[H_0(t-t_0)]^n \label{eqs-3},
\end{equation}
\noindent
where
\begin{equation}
A_n=\frac{a^{(n)}}{aH^n} \label{eqs-4},
\end{equation}

where $a^{(n)}=\frac{d^na}{dt^n}$. For $\Lambda$CDM model in a FRLW background, we have
\begin{eqnarray}
  A_2&=&1-\frac{3}{2}\Omega_m \label{eqs-5}, \\
  A_3&=&1 \label{eqs-6},  \\
  A_4&=&1-\frac{3^2}{2}\Omega_m \label{eqs-7},  \\
  A_5&=&1+3\Omega_m+\frac{3^3}{2}\Omega^2_m  \label{eqs-8}.
\end{eqnarray}

For $\Lambda$CDM model, we have $\Omega_m=\frac{2}{3}(1+q)$. Statefinder hierarchy $S_n$ is defined as~\cite{state17}
\begin{eqnarray}
  S_2&=&A_2+\frac{3}{2}\Omega_m \label{eqs-9}, \\
  S_3&=&A_3 \label{eqs-10},   \\
  S_4&=&A_4+\frac{3^2}{2}\Omega_m \label{eqs-11},   \\
  S_5&=&A_5-3\Omega_m-\frac{3^3}{2}\Omega^2_m  \label{eqs-12},...
\end{eqnarray}

Above equations define a null test diagnostic for $\Lambda$CDM~\cite{state17},
\begin{equation}
S_n\mid_{\Lambda CDM}=1 \label{eqs-12'}.
\end{equation}

When $n\geq3$, one can define a new null test
diagnostic for $\Lambda$CDM~\cite{state17}
\begin{eqnarray}
  S^{(1)}_3&=&S_3 \label{eqs-13}, \\
  S^{(1)}_4&=&A_4+3(1+q) \label{eqs-14},   \\
  S^{(1)}_5&=&A_5-2(4+3q)(1+q)  \label{eqs-15},...
\end{eqnarray}

\noindent
This series of Statefinders for $\Lambda$CDM are invariable during the evolution of
the universe~\cite{state17}
\begin{equation}
S^{(1)}_n\mid_{\Lambda CDM}=1 \label{eqs-16}.
\end{equation}

One can construct a second member of the Statefinders on the basis of $S^{(1)}_n$~\cite{state17}:
\begin{equation}
S^{(2)}_n=\frac{S^{(1)}_n-1}{3(q-\frac{1}{2})} \label{eqs-17}.
\end{equation}

For $\Lambda$CDM, we have~\cite{state17}
\begin{equation}
S^{(2)}_n\mid_{\Lambda CDM}=0 \label{eqs-18}.
\end{equation}

So corresponding to the diagnostic plains $S_n-S^{(1)}_n$, $S_n-S^{(2)}_n$, $S^{(1)}_n-S^{(2)}_n$, for $\Lambda$CDM one have $\{S_n,S^{(1)}_n \}=\{1,1 \}$, $\{S_n,S^{(2)}_n \}=\{1,0 \}$, $\{S^{(1)}_n,S^{(2)}_n \}=\{1,0 \}$. For dynamical dark energy models, Eqs.~(\ref{eqs-16})and ~(\ref{eqs-18}) will be broken. Now we will use the Statefinder hierarchy to discriminate our two parametric models from $\Lambda$CDM. Write down the deceleration parameter$q$, $A_3$, $A_4$, $A_5$
\begin{eqnarray}
  q&=&(1+z)\frac{1}{E}\frac{dE}{dz}-1 \label{eqs-19}, \\
  A_3&=&(1+z)\frac{1}{E^2}\frac{d[E^2(1+q)]}{dz}-3q-2 \label{eqs-20},   \\
  A_4&=&-(1+z)\frac{1}{E^3}\frac{d[E^3(2+3q+A_3)]}{dz}+4A_3+3q(q+4)+6 \label{eqs-21},   \\
  A_5&=&-(1+z)\frac{1}{E^4}\frac{d[E^4(A_4-4A_3-3q(q+4)-6)]}{dz}+5A_4-10A_3(q+2)-30q(q+2)-24.  \label{eqs-22}
\end{eqnarray}

\begin{figure}[h!]
\begin{center}
\includegraphics[width=0.5\textwidth]{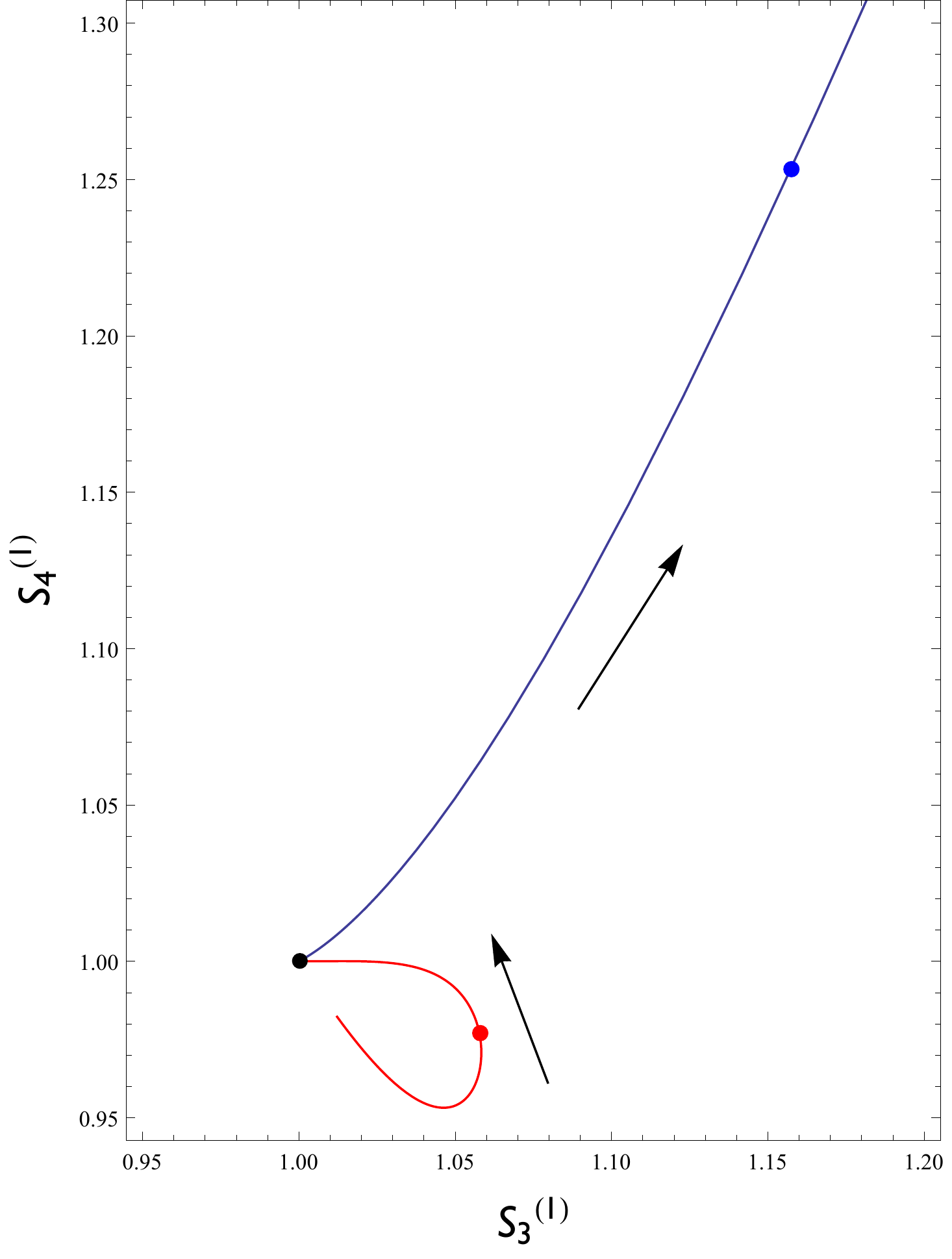}
\end{center}
\caption{The Statefinder $\{S^{(1)}_3,S^{(1)}_4 \}$ are shown for model 1, model 2 and $\Lambda$CDM model. The trajectories of model 1 and model 2 correspond to red line and blue line, respectively. The fixed point $\{1,1 \}$ represents the $\Lambda$CDM model. The present epoch in different models is shown as a dot and the arrows indicate the evolutional direction with respect to time.}
\label{fig:statefinder1}
\end{figure}

\begin{figure}[h!]
\begin{center}
\includegraphics[width=0.5\textwidth]{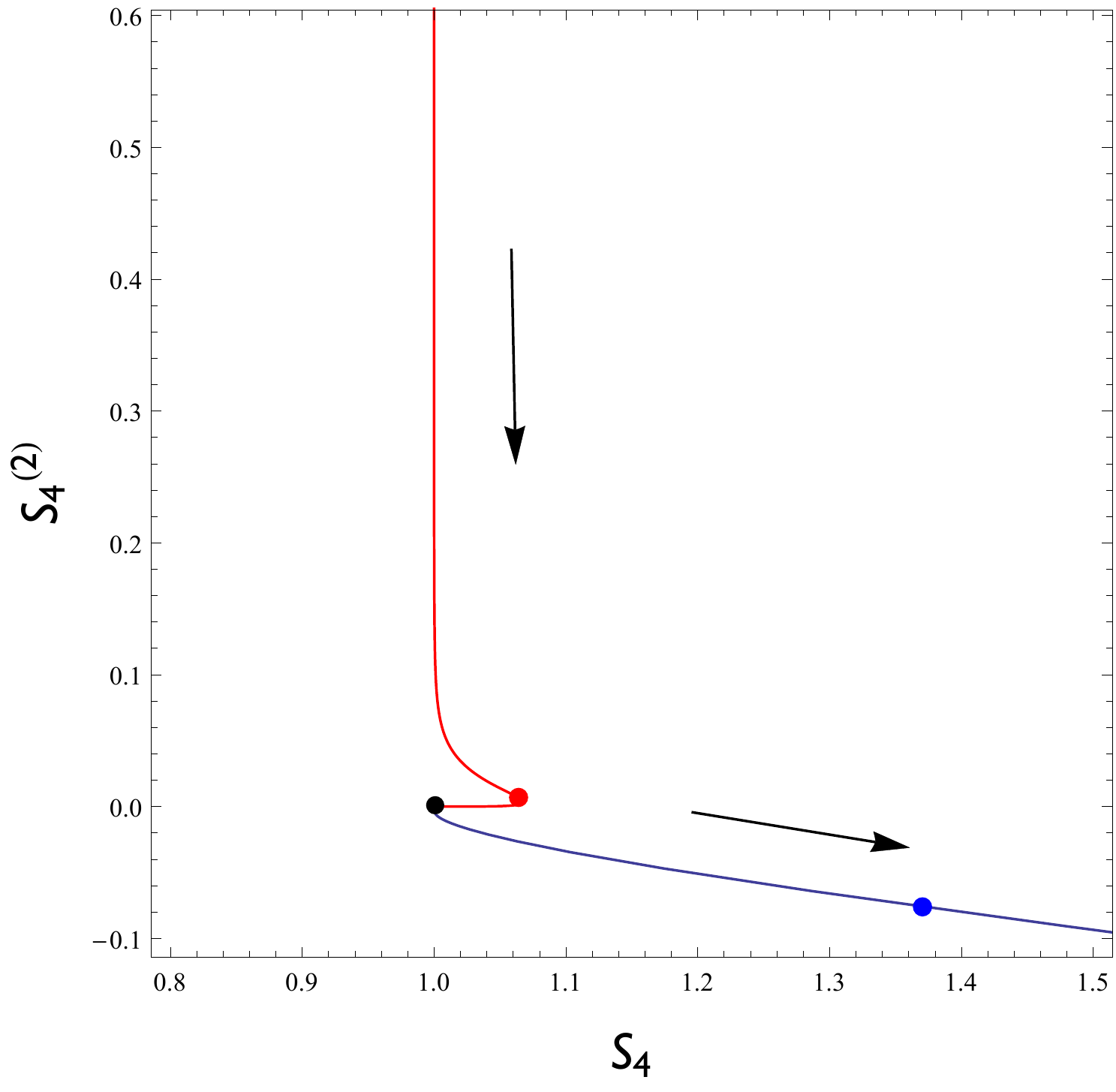}
\end{center}
\caption{The Statefinder $\{S_4,S^{(2)}_4 \}$ are shown for model 1, model 2 and $\Lambda$CDM model. The trajectories of model 1 and model 2 correspond to red line and blue line, respectively. The fixed point $\{1,1 \}$ represents the $\Lambda$CDM model. The present epoch in different models is shown as a dot and the arrows indicate the evolutional direction with respect to time.}
\label{fig:statefinder2}
\end{figure}

\begin{figure}[h!]
\begin{center}
\includegraphics[width=0.5\textwidth]{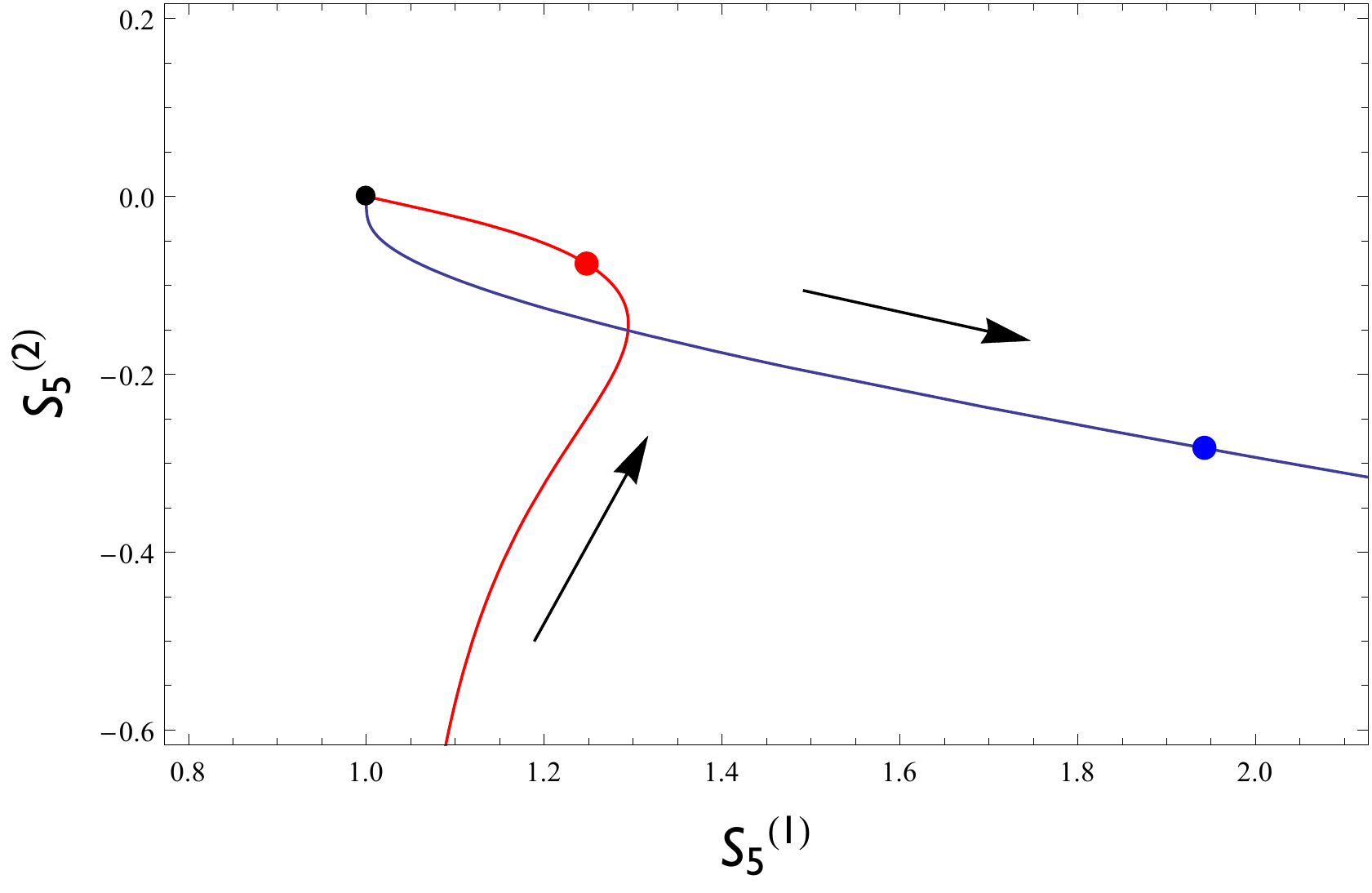}
\end{center}
\caption{The Statefinder $\{S^{(1)}_5,S^{(2)}_5 \}$ are shown for model 1, model 2 and $\Lambda$CDM model. The trajectories of model 1 and model 2 correspond to red line and blue line, respectively. The fixed point $\{1,1 \}$ represents the $\Lambda$CDM model. The present epoch in different models is shown as a dot and the arrows indicate the evolutional direction with respect to time.}
\label{fig:statefinder3}
\end{figure}

$\{S^{(1)}_3,S^{(1)}_4 \}$ is an often-used Statefinder, in~\cite{state17} it was used for distinguishing CG, DPG, $\omega$CDM and $\Lambda$CDM, also in~\cite{state18} it was used for distinguishing GCG, MCG, SCG, PKK and $\Lambda$CDM. Using the diagnostic $\{S^{(1)}_3,S^{(1)}_4 \}$ to distinguish two parametric models and $\Lambda$CDM, as shown in Fig.~(\ref{fig:statefinder1}), we can find that model 1 and model 2 can be well distinguished from $\Lambda$CDM at the present epoch, consequently and in the future. Moreover, it is worth noting that, at the present epoch model 1 and model 2 correspond to the points $\{1.058, 0.977\}$ and $\{1.157, 1.253\}$, respectively. Furthermore, one can obtain the conclusion that model 1 will evolve into a de-sitter universe in the end and model 2 will gradually deviate from the stand cosmological model.

In the plains of Statefinder $\{S_4,S^{(2)}_4 \}$ and $\{S^{(1)}_5,S^{(2)}_5 \}$ , one can easily find that the two models are both well distinguishable from $\Lambda$CDM. At the same time, we can get the same result as the Statefinder $\{S^{(1)}_3,S^{(1)}_4 \}$ that the evolutional trajectories of model 1 and model 2 can be well distinguished. Interestingly, one could discover that in this case, the trajectories of two models will have an unexpected overlap, which means that the universes corresponding to the two models will share the same phase at different times in the past.

\subsection{The growth rate of perturbations}
In the following context, we adopt the fractional growth parameter as a supplement for the statefinders \cite{1,2}
\begin{equation}
\epsilon(z)=\frac{f(z)}{f_{\Lambda CDM}(z)},
\end{equation}
where $f(z)\approx\Omega_m(z)^\gamma$ reflects the growth rate of linearized density perturbations \cite{3} and
\begin{equation}
\gamma(z)=\frac{3}{5-\frac{\omega}{1-\omega}}+\frac{3}{125}\frac{(1-\omega)(1-1.5\omega)}{(1-1.2\omega)^3}[1-\Omega_m(z)]+\mathcal{O}[(1-\Omega_m(z))]^2.
\end{equation}
In general, one can only calculate the former two terms to obtain a good approximation for a concrete physical dark energy model. For an instance, $\gamma\approx0.55$ in the standard cosmological model \cite{3,5}. However, in the case of extended theories of gravity (ETG), the situation will be different since in ETG, the perturbation growth contains information which is complementary to that contained in the expansion history (see Ref. \cite{4}).

\begin{figure}
\centering
\includegraphics[scale=0.5]{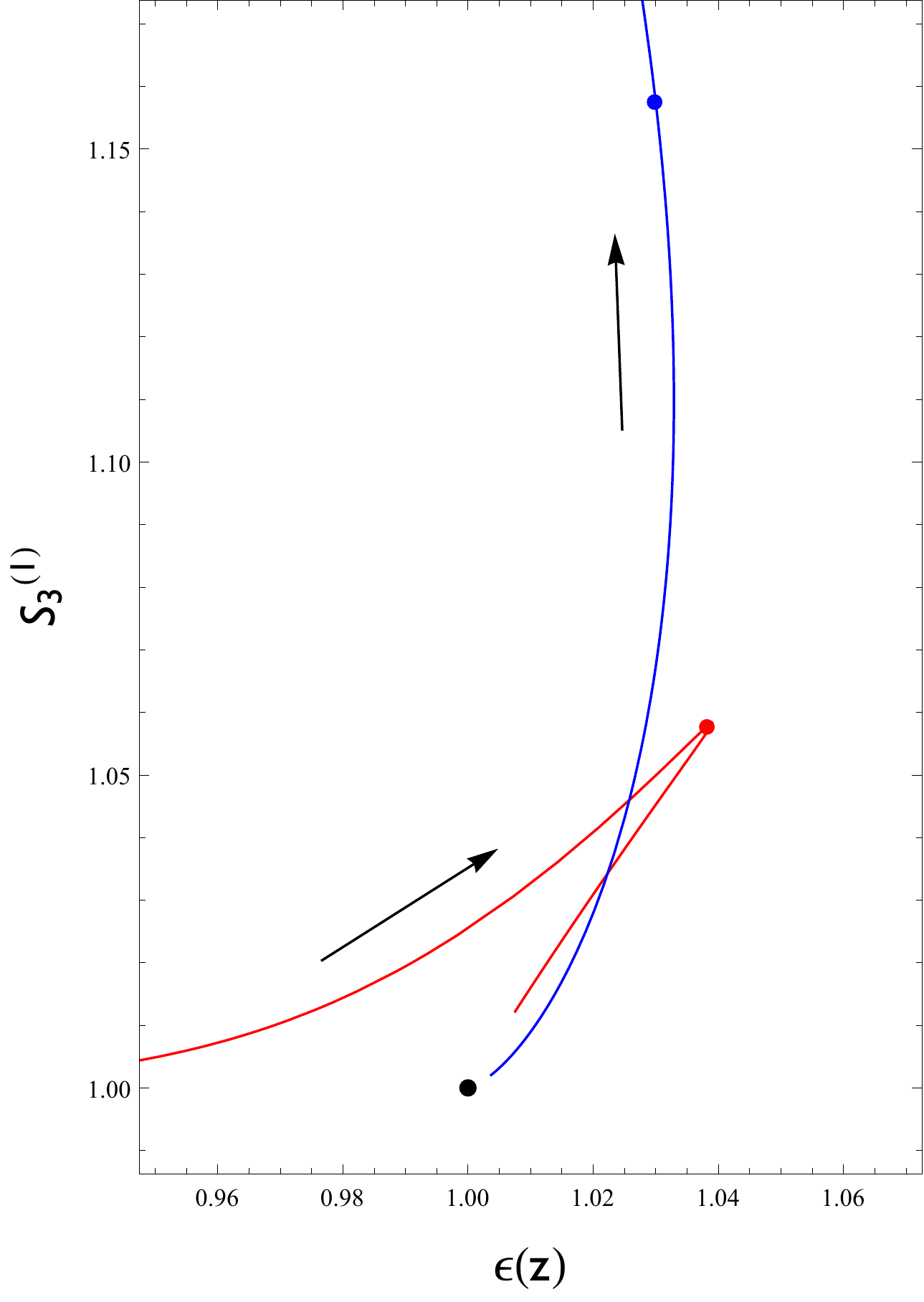}

\caption {The simply composite null diagnostic $\{\epsilon(z),S_3^{(1)}\}$ is shown for model 1, model 2 and $\Lambda$CDM model. The fixed point \{1,1\} represents the $\Lambda$CDM model and the arrows indicate the evolutional direction with respect to time. The trajectories of model 1 and model 2 correspond to red line and blue line, respectively. One can easily discover that the two models can be well distinguished at the low redshift, consequently at the present epoch and in the far future.}\label{fig5}
\end{figure}

For the above-mentioned reason, combining the statefinders with the fractional growth parameter, one can conveniently define a simply composite null diagnostic (SCND), namely, $\{\epsilon(z),S_3^{(1)}\}$, where the fixed point \{1,1\} corresponds to the $\Lambda$CDM model. From Fig. (\ref{fig5}), one can apparently discover that model 1 and model 2 are well distinguished by the SCND at the present epoch, consequently and in the far future. Similarly, we can see that from Fig. {\ref{fig5}}, the trajectory of model 1 share two same points with that of model 2 at different times, which implies that there appear to be a more fundamental model based our previous models.

\section{rip analysis}
From the conservation equation $\dot{\rho}=-3H(\rho+P)$, we know in the physical scenario of phantom field the EoS of dark energy $\omega_{de}<-1$, which means that the energy density will increase in the future. Subsequently~\cite{rip}, the possible fates of the universe can be divided into several categories based on the time asymptotics of the Hubble parameter $H(t)$: the big rip, for which $H(t)\rightarrow\infty$ at finite time; the little rip, for which $H(t)\rightarrow\infty$ as time goes to infinity; and the pseudo-rip, for which $H(t)\rightarrow constant$, as time goes to infinity which is an intermediate case between the de-Sitter cosmology and the little rip. In this section we will show the fates of model 1 and model 2 in the physical scenario of phantom field.

For model 1, substitute Eq.~(\ref{eq2-4}) to the Friedmann equation
\begin{equation}
 H^2=\frac{8\pi G}{3}\rho_0[\alpha+\beta a^{-1}+(1-\alpha-\beta)a^{-3}] \label{eqrip-1}.
\end{equation}

In the above equation, we can find in the future term ``$\frac{8\pi G}{3}\rho_0\alpha$'' will dominate the
right hand side of the equation, that means $H(t)\rightarrow constant$  as time goes to infinity; namely, there exists a pseudo-rip for model 1 in the future.

For model 2, substitute Eq.~(\ref{eq2-10}) to the Friedmann equation
\begin{equation}
 H^2=\frac{8\pi G}{3}\rho_0[\gamma+\delta a+(1-\gamma-\delta)a^{-3}] \label{eqrip-2}.
\end{equation}

In above equation, we can find in the future term``$\frac{8\pi G}{3}\rho_0\delta a$'' will dominate the right side of the equation in the case of $\delta>0$, so Eq.~(\ref{eqrip-2}) can be simplified as
\begin{equation}
 \frac{\dot{a}}{a}=(\frac{8\pi G\rho_0\delta}{3})^{\frac{1}{2}}a^{\frac{1}{2}} \label{eqrip-3}.
\end{equation}

Solve the differential equation~(\ref{eqrip-3}), we can obtain the scale factor $a$ as a function of time $t$
\begin{equation}
 a=\frac{4}{[2-N(t-t_0)]^2} \label{eqrip-4},
\end{equation}

with
\begin{equation}
 N=(\frac{8\pi G\rho_0\delta}{3})^\frac{1}{2} \label{eqrip-5},
\end{equation}

\noindent
where $t_0$ is the present value of time. Substitute Eq.~(\ref{eqrip-4}) to Eq.~(\ref{eqrip-3}), we get

\begin{equation}
 H(t)=\frac{2N}{2-N(t-t_0)} \label{eqrip-6}.
\end{equation}

Note at a finite time $t_{rip}=\frac{2}{N}+t_0$, the Hubble parameter $H\rightarrow\infty$, this is to say the universe runs into a big rip. Here we choose $t_0=0$,so the finite time $t_{rip}=\frac{2}{N}$. Using Friedmann equation $H_0=(\frac{8\pi G}{3}\rho_0)^\frac{1}{2}$, the Eq.~(\ref{eqrip-5}) can be written as $N=H_0(\delta)^\frac{1}{2}$, we get
\begin{equation}
 t_{rip}=\frac{2}{H_0(\delta)^\frac{1}{2}}\label{eqrip-7},
\end{equation}
where $H_0=67.8$ from Recent observations. It is easy to see the finite time just relate to  parameter $\delta$. We can read from~\cite{manuscript} that $\sigma$ standard deviation of $\delta$ is 0.120. In the scenario of phantom, $\delta$ can take the value from 0 to 0.199. Due to the propagation of error, we get $t_{rip}=1.028^{+0.784}_{-0.514}\times10^{11}$. This means that $t_{rip}$ will shift significantly even if $\delta$ only varies a little. By the previous deduction, our universe will terminate after $10^{11}$ years.

\section{Conclusions}
In this paper, at first, we have used the $Om$ diagnostic to discriminate two parametric models for the total pressure and $\Lambda$CDM model, and get the conclusion that the two models can be hardly distinguished from each other as well as the $\Lambda$CDM model in terms of 68\% confidence range.

In the second place, we have applied the Statefinder hierarchy and the growth rate of perturbations into discriminating the two models with the $\Lambda$CDM model. The results indicate that our two parametric models can be distinguished from the $\Lambda$CDM model, and in comparison to the $Om$ diagnostic, model 1 and model 2 can be distinguished from each other better. Interestingly, we discover that through these three geometrical diagnostics, there may exist a more fundamental model based our previous models.

At last, we investigate the fates of universe evolution of our two models by means of the rip analysis. In model 1, we find that there exits a pseudo-rip in the future, which means the behavior of model 1 tends to a de-sitter universe in the future. In model 2, we find that the universe will runs into a big rip at a finite time $t=t_{rip}=1.028^{+0.784}_{-0.514}\times10^{11}$ in the scenario of phantom, we guess that with the growth of the energy density near to rip the effects of quantum gravity (string/M-theory) may become dominant to keep away from the doomsday. The attractive work in the future could be to consider the astrophysical scales effects of our models, by assuming that the dark energy is permeated everywhere in the universe.

\section*{Acknowledgements}
We are grateful for Jiaxin Wang's precious suggestions and help. G. Y. and D. W. thank Qixiang Zou and Qiang Zhang for useful discussions. This work is partly supported by the the National Science Foundation of China.


\begin{thebibliography}{99}
\bibitem{a1}
Riess A.G.et al,
\newblock{\em Astro. J.} {\bf 116,} 1009 (1998).

\bibitem{a2}
Perlmutter S.J. et al,
\newblock{\em Astro. J.} {\bf 517,} 565 (1999).


\bibitem{0}
P. A. R. Ade et al., [Planck Collaboration],  arXiv:1502.01589v2 [astro-ph.CO]

\bibitem{manuscript}
Qiang Zhang, Guang Yang, Qixiang Zou, Xin-he Meng and Ke-ji Shen. Exploring the low redshift universe: two parametric models
for effective pressure, Eur. Phys. J. C {\bf 75}, 300 (2015).




\bibitem{state19}
V. Sahni, A. Shafieloo, and A. A. Starobinsky, Two new diagnostics of dark energy, Phys. Rev. D {\bf 78}, 103502 (2008) [arXiv:0807.3548].

\bibitem{state1}
V. Sahni, T. D. Saini, A. A. Starobinsky and U. Alam  Statefinder¡ªA new geometrical diagnostic of dark energy, JETP Lett. {\bf 85}, 201 (2003) [astro-ph/0201498].

\bibitem{state1'}
U. Alam, V. Sahni, T. D. Saini and A. A. Starobinsky, Exploring the expanding Universe and dark energy using the
statefinder diagnostic, Mon. Not. Roy. Astron. Soc. {\bf 344}, 1057 (2003) [astro-ph/0303009].


\bibitem{state17}
M. Arabsalmani and V. Sahni, Statefinder hierarchy: An extended null diagnostic for concordance cosmology, Phys. Rev.D {\bf 83}, 043501 (2011) [arXiv:1101.3436].

\bibitem{1}
V. Acquaviva et al, Next generation redshift surveys and the origin of cosmic acceleration, Phys. Rev. D {\bf 78}, 043514 (2008).

\bibitem{2}
V. Acquaviva et al, How to falsify the GR + $\Lambda$CDM model with galaxy redshift surveys, Phys. Rev. D {\bf 82}, 082001 (2008).


\bibitem{state10}
X.-T. Gao and R.-J. Yang, Geometrical diagnostic for purely kinetic k-essence dark energy, Phys. Lett. B {\bf 687}, 99 (2010) [arXiv:1003.2786].

\bibitem{state11}
L. N. Granda, W. Cardona, and A. Oliveros, Current observational constraints on holographic dark energy model, [arXiv:0910.0778v1].

\bibitem{state16}
R. Yang, J. Qi, and B. Chen, Discriminate spatial Ricci scalar dark energy from $\Lambda$CDM, Sci China-Phys Mech Astron {\bf 55},  1952 (2012).


\bibitem{state2}
S. Chongchitnan and G. Efstathiou  Can we ever distinguish between quintessence and a cosmological constant? Phys.Rev. D {\bf 76}, 043508 (2007) [arXiv:0705.1955].

\bibitem{state3}
Eric V. Linder. The dynamics of quintessence, the quintessence of dynamics, Gen. Rel. Grav. {\bf 40}, 329 (2008)  [arXiv:0704.2064].

\bibitem{state4}
V. Gorini, A. Kamenshchik, and U. Moschella, Can the Chaplygin gas be a plausible model for dark energy? Phys.\ Rev.\ D  {\bf 67}, 063509 (2003) [astro-ph/0209395].

\bibitem{state5}
W. Chakraborty, U. Debnath, and S. Chakraborty, Generalized cosmic Chaplygin gas model with or without interaction, Grav. Cosmol. {\bf 13}, 294 (2007) [arXiv:0711.0079].

\bibitem{state6}
S. Li, Y. Ma, Y. Chen, Dynamical Evolution of Interacting Modified Chaplygin Gas, Int. J. Mod. Phys. D  {\bf 18}, 1785 (2009) [arXiv:0809.0617].

\bibitem{state7}
G. Panotopoulos, Statefinder parameters in two dark energy models, Nucl. Phys. B  {\bf 796}, 66 (2008) [arXiv:0712.1177].

\bibitem{state8}
R. Myrzakulov, M. Shahalam, Statefinder hierarchy of bimetric and galileon models for concordance cosmology, JCAP {\bf 10}, 047 (2013) [arXiv:1303.0194].

\bibitem{state9}
M. Sami, M. Shahalam, M. Skugoreva, and A. Toporensky, Cosmological dynamics of a nonminimally coupled scalar field system and its late time cosmic relevance, Phys. Rev. D {\bf 86}, 103532 (2012) [arXiv:1207.6691].



\bibitem{state12}
J. Zhang, X. Zhang, H. Liu, Statefinder diagnosis for the interacting model of holographic dark energy, Phys. Lett. B {\bf 659}, 26 (2008) [arXiv:0705.4145].

\bibitem{state13}
C.-J. Feng, Statefinder diagnosis for Ricci dark energy, Phys. Lett. B {\bf 670}, 231 (2008) [arXiv:0809.2502].

\bibitem{state14}
H. Wei and R.-G. Cai, Statefinder diagnostic and $\omega-\omega^\prime$ analysis for the agegraphic dark energy models without and with interaction, Phys. Lett. B {\bf 655}, 1 (2007) [arXiv:0707.4526].

\bibitem{state15}
P. Wu and H. Yu, Statefinder Parameters for Quintom Dark Energy Model, Int. J. Mod. Phys. D {\bf 14}, 1873 (2005) [gr-qc/0509036].



\bibitem{state18}
Jun Li, Rong-Jia Yang, Bohai Chen, Discriminating dark energy models by using the statefinder hierarchy and the growth rate of matter perturbations, JCAP {\bf 12}, 043 (2014) [arXiv:1406.7514v2].

\bibitem{3}
L. Wang and P. J. Steinhardt, Cluster Abundance Constraints on Quintessence Models, Astrophys. J. {\bf 508}, 483 (1998).

\bibitem{4}
Maryam Arabsamani and Varun Sahni, Statefinder hierarchy: An extended null diagnostic for concordance cosmology, Phys. Rev. D {\bf 83}, 043501 (2011).

\bibitem{5}
E. V. Linder, Cosmic growth history and expansion history, Phys. Rev. D {\bf 72}, 043529 (2005).

\bibitem{rip}
P.~H.~Frampton, K.~J.~ Ludwick and R.~J.~Scherrer, Pseudo-rip: Cosmological models intermediate between the cosmological constant and the little rip, Phys.\ Rev.\ D {\bf 85}, 083001 (2012) [arXiv:1112.2964v2].


\end{thebibliography}
\end{document}